\documentclass{easychair}

\usepackage{doc}
\usepackage{graphicx}
\usepackage{subcaption}
\usepackage{multirow}
\usepackage{incgraph,tikz}
\usepackage[T1]{fontenc}

\title{A Systematization of Cybersecurity Regulations, Standards and Guidelines for the Healthcare Sector}

\author{
Maria Patrizia Carello\inst{1}
\and
    Alberto Marchetti Spaccamela\inst{1}
\and
   Leonardo Querzoni\inst{1}
   \and
Marco Angelini\inst{1}
}

\institute{
  Sapienza University of Rome\\
  \email{\{carello, alberto, querzoni, angelini\}@diag.uniroma1.it}\\
 }

\authorrunning{Carello, Marchetti, Querzoni and Angelini}

\titlerunning{Systematization of Cybersecurity doc. for the Health Sector}

\begin{document}

\maketitle

\begin{abstract}

The growing adoption of IT solutions in the healthcare sector is leading to a steady increase in the number of cybersecurity incidents. As a result, organizations worldwide have introduced regulations, standards, and best practices to address cybersecurity and data protection issues in this sector. However, the application of this large corpus of documents presents operational difficulties, and operators continue to lag behind in resilience to cyber attacks. This paper contributes a systematization of the significant cybersecurity documents relevant to the healthcare sector. We collected the 49 most significant documents and used the NIST cybersecurity framework to categorize key information and support the implementation of cybersecurity measures.

\end{abstract}



\section{Introduction}
\label{sec:intro}

Worldwide, the \textit{digital transformation of health services} is seen as an important and influential process, increasing the integration of technology in healthcare organizations, ranging from the use of computers and electronic health records to home monitoring of patients, electronic medical devices, and decision support systems \cite{ricciardi2019govern}.\\
Digital transformation affects many aspects of healthcare systems and allows for the improvement of service quality. For example, it is known that the adoption of telemedicine decreases hospital mortality rates without a significant increase in cost \cite{Armaignac2018ImpactOT,haleem2021telemedicine}.\\
However, the extensive integration of technologies into existing organizations has caused cybersecurity incidents to become an increasing challenge. Therefore, preventing, mitigating, responding to, managing emergencies, and recovering from cyber-attacks are critical responsibilities in the health domain nowadays.\\
To answer the above needs, several regulations, standards, and best practices on healthcare security have been proposed worldwide to help and guide health organizations in improving their cybersecurity preparedness.
However, the correct application of regulations, standards, and best practices poses several issues. Firstly, these guidelines have been designed by different actors for various purposes, and their fragmented nature makes integration and application challenging (\textit{issue1}). Furthermore, they often provide a high-level overview of security measures in a discursive manner without specifying the technical security policies that need to be implemented (\textit{issue2}). Moreover, there is significant overlap among documents published by different sources, and different terminology is used to refer to the same concepts (\textit{issue3}). Finally, the extensive use of legal jargon and cross-references to other regulations makes it difficult to parse and extract security-focused elements (\textit{issue4}).\\
This paper proposes a systematization of the corpus of documents mentioned above to overcome these issues. We extract succinct and informative excerpts related to security and data protection from non-technical sources, and then provide a consistent view of the stated security measures by analyzing the degree of overlap and filling the gaps in coverage of security-related aspects.
To accomplish this, we began by analyzing a corpus of 68 documents to identify relevant ones. From these, we extracted excerpts of interest and mapped them to the NIST Cybersecurity Framework \cite{NISTFR}. Based on each mapped excerpt, we defined a set of cybersecurity controls that can be effectively used to build cybersecurity plans.\\
We also present the methodology used to conduct our study and exemplify its application in the healthcare sector, discussing findings that highlight possible areas for improvement.\\
The paper is organized as follows: Section \ref{sec:related} provides background information on cybersecurity regulations, standards, and best practices issued worldwide in the healthcare sector and illustrates related proposals; Section \ref{sec:methodology} introduces our novel methodology for systematizing such corpus of documents, and describes its results; Section \ref{sec:findings} discusses important findings identified through this systematization; Section \ref{sec:conclusions} concludes the paper.
\section{Background and Related Work}
\label{sec:related}
The corpus of documents that govern cybersecurity and data protection for healthcare organizations can be grouped into three categories: \emph{Regulations}, \emph{Standards}, and \emph{Best Practices}. Each category is briefly described in the following.\\

\noindent
{\bf Regulations} are issued by an executive authority or regulatory agency and have the force of law. They can be national or international (for the national ones, in this paper we refer to the Italian regulations). One of the first security regulations for the healthcare sector is the U.S. \emph{Health Insurance Portability and Accountability Act (HIPAA)}~\cite{HIPAA}, 1996. The main goal of HIPAA was to protect Personally Identifiable Information (PII) and to preserve privacy while allowing individuals to access their medical records. HIPAA was updated in 2003 and 2013, adding requirements for managing Electronic Protected Health Information and implementing penalties for privacy  violations.
The EU \emph{General Data Protection Regulation (GDPR)}~\cite{GDPR}, 2018, 
regulates the processing and circulation of personal data;
GDPR recognizes health data as special data that requires greater protection and specific security measures.
 The European Union issued the \emph{Regulation on Medical Devices (MDR)}, 2017, that presents cybersecurity requirements of medical devices~\cite{MDR}.\\

\noindent
{\bf Standards} are documents set up by authority or general consent as a model or example to be compliant with. In the last few years, several Standards have been released to promote the development of security requirements for the healthcare sector, for example, the \emph{ISO 27799 Health informatics} 
~\cite{27799}, 2016, provides an implementation guide for the controls described in ISO/IEC 27002 and supplements them where necessary. 
More recently, the \emph{ISO/TR 21332 - Health informatics} 
~\cite{21332}, 2021, provides an overview of the security and privacy of Electronic Health Records (EHR) in a cloud computing service and the
\emph{IEC 80001-1} 
~\cite{80001}, 2021, specifies security requirements 
for connecting  medical devices.\\

\noindent
{\bf Best Practices}
 are guidelines  to be used in a particular business or industry (such as healthcare) to meet cybersecurity objectives and to be compliant with regulations. For example, the \emph{NIST Security Rule -SP 800-66}~\cite{NISTSP}, 2008, summarizes HIPAA security standards to support healthcare organizations to be compliant with HIPAA regulations.
In Europe, ENISA published several documents; we mention the \emph{Procurement guidelines for cybersecurity in hospitals}, 2020, 
~\cite{ProcurementGuidelines} and the  \emph{European Commission's (EC) Medical Devices Coordination Group (MDCG)} published in 2020 a guide on how to fulfill all the essential cybersecurity requirements issued by the MDR and IVDR (In Vitro Diagnostic Medical Devices Regulation) regulations~\cite{MDCG}. \\
\noindent
{{\bf Related work.}}
In the last ten years there has been a significant increase  in the pace of publication about cybersecurity and healthcare  ~\cite{argaw2020cybersecurity}.
The interest is motivated by  the  key role of Cybersecurity in the healthcare sector:  any disruption in health services can be a disaster for patients' health, not only  for organizations. In the following, we focus on studies that aim to provide a systematization for the large number of cybersecurity documents in the healthcare domain. \\
Jalali et al.~\cite{Jalali2019} conducted a broad work on scientific literature: they surveyed 472 scientific contributions extracted from Pubmed and Web of Science at the intersection of cybersecurity and healthcare. Their findings show that most contributions focus on technological aspects, while 32\% focus on managerial and policy-making topics. Differently from our work, they do not consider regulations, standards, and best practices, making their work complementary to our approach.
Mohammed~\cite{mohammed2017us} discusses the compliance issues and challenges for healthcare organizations in the U.S., focusing on  HITECH and HIPAA.
The author lists among major challenges the vagueness and ambiguity of many of the prescriptions of those documents, similar to what we identified before. 

Furthermore, it has been observed how cybersecurity standards and regulations are still uncertain, overlapping, and do not entirely address healthcare-specific concerns; as a consequence complying with cybersecurity rules is a challenging activity that involves time and expense for healthcare organizations, hindering their ability to develop adequate cybersecurity programs \cite{coventry2018cybersecurity,akinsanya2019current,biasin2022cybersecurity_AI,martin2017cybersecurity,burke2019cybersecurity}.
In~\cite{lechner2017overview} and ~\cite{THOMASIAN2021100549} regulations and standards for medical device software are considered focusing  on the device manufacturer as the intended target; our goal is to inform the management (i.e., CISOs and DPOs) inside healthcare organizations.

As a consequence of the aforementioned considerations, it is necessary to support healthcare organizations in navigating and making sense of these documents to support the extraction and modeling of cybersecurity measures.
\section{Methodology}
\label{sec:methodology}
This section outlines a novel four-step methodology for the systematization of cybersecurity regulations, standards, and best practices that have been published over time for the healthcare sector. In the first step, we thoroughly searched public repositories to find documents of interest. In the second step, the documents are analyzed to identify excerpts that refer to technical security and governance measures. In the third step, cybersecurity excerpts are mapped on the Subcategories of the NIST Framework \cite{NISTFR}, and in the last step, a control definition procedure is carried out for each subcategory. \\
All the results and additional materials are available at this link:\\ \url{https://github.com/carelloSapienza/Systematization-healthcare}.

\begin{figure}[h]
\centering \includegraphics[width=0.9\textwidth]{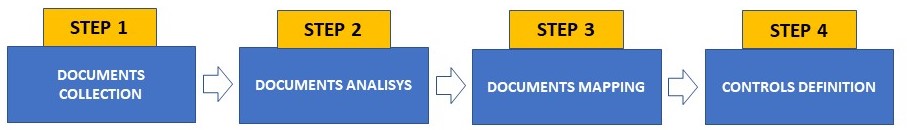}
\caption{Methodology Steps}
\label{fig:methodology}
\end{figure}

\subsection{Documents Collection}
\label{subsec:docCollection}

We explored the information available on the main official sources of European, International, and National regulators (e.g., ENISA, NIST, Salute.gov) 
using main searched keywords such as \textit{cybersecurity, privacy, electronic health record (EHR), medical device, telemedicine, cloud for healthcare}.

A second round of research was conducted using the main indexing platforms (e.g., Elsevier Scopus, Google scholar, IEEE Xplore), and the primary searched keywords were: \textit{cybersecurity in healthcare, healthcare cybersecurity legislation, telemedicine security and privacy, cybersecurity of medical devices; security framework for healthcare}. We performed a forward and backward analysis for each document or paper collected.
Afterward, to refine the research, we constrained each collected document to two key requirements:
\emph{(i)} the regulation must be in effect, \emph{(ii)} the document must address data security, privacy issues, or cybersecurity measures for healthcare organizations or public administrations.
Therefore, we did not include works that address only manufacturers of medical devices, external service providers, or government agencies.\\
Examples of documents excluded, as not deemed of interest, are \emph{ISO/TR 17522:2015 Health informatics — Provisions for health applications on mobile/smart devices}~\cite{17522}, focused only on interoperability, and \emph{ISO 14971:2019 Medical devices — Application of risk management to medical devices}~\cite{14971} that is specifically addressed to manufacturers of medical devices.

\medskip
\noindent
\textbf{Results.}
This step allowed us to gather 68 potential documents of interest first, then narrowed to 49 documents by considering the key requirements.
The final corpus, therefore, is composed of \emph{11 regulations}, \emph{21 best practices}, and \emph{17 standards} gathered by European (\emph{9}), international (\emph{19}), and national (\emph{21}) sources.

\subsection{Documents Analysis}
\label{subsec:docAnalysis}

In this step, each of the 49 documents previously collected is accurately analyzed to identify key excerpts of text that refer to technical security and governance measures.
A key excerpt of text is a sentence in a document that refers to areas of cybersecurity or data protection, such as information security policies, data privacy, incident management, etc.
The identification has been performed manually by at least two members of our team with expertise in security governance, cybersecurity, and data protection.
Once identified, the excerpt is extracted from its original document and collected in a table as output for the next step.
Another group of information security specialists has regularly examined the collected excerpts to verify their relevance.\\
This step mitigates \emph{issue 4} helping to organize the texts and to extract only the relevant contents (security and data protection).

\begin{figure}[ht]
\centering{\includegraphics[width=0.95\textwidth]{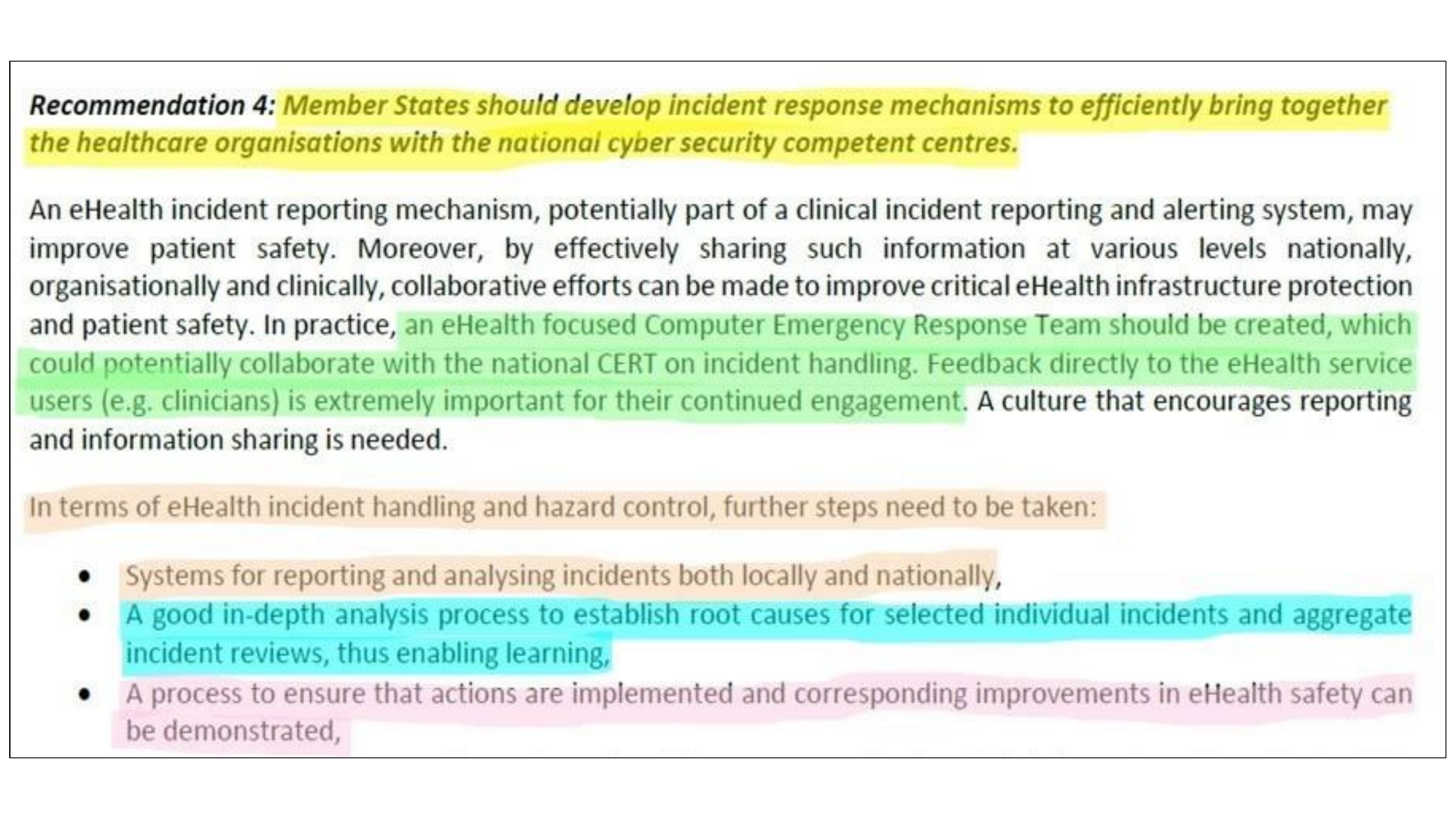}}
\caption{Example of Key Excerpts Extraction from \cite{ENISAsec} }
\label{fig:contentExt}
\end{figure} 

\medskip
\noindent
\textbf{Results.}
Figure~\ref{fig:contentExt} shows an example of key excerpts identification on the document \textit{Security and Resilience in eHealth Infrastructures and Services}~\cite{ENISAsec}.\\
To identify relevant key excerpts from non-relevant ones, consider the first sentence: \emph{``An eHealth incident reporting mechanism, potentially part of a clinical incident reporting and alerting system, may improve patient safety''}.
This excerpt does not give any technical information and therefore is not a key excerpt.
Conversely, the sentence highlighted in green asserting that \emph{``Computer Emergency Response Team should be created''} and \emph{``could potentially collaborate with the national CERT''} has been considered a relevant excerpt since it provides clear cybersecurity indications.

\begin{figure}[h]
\centering \includegraphics[width=0.4\textwidth]{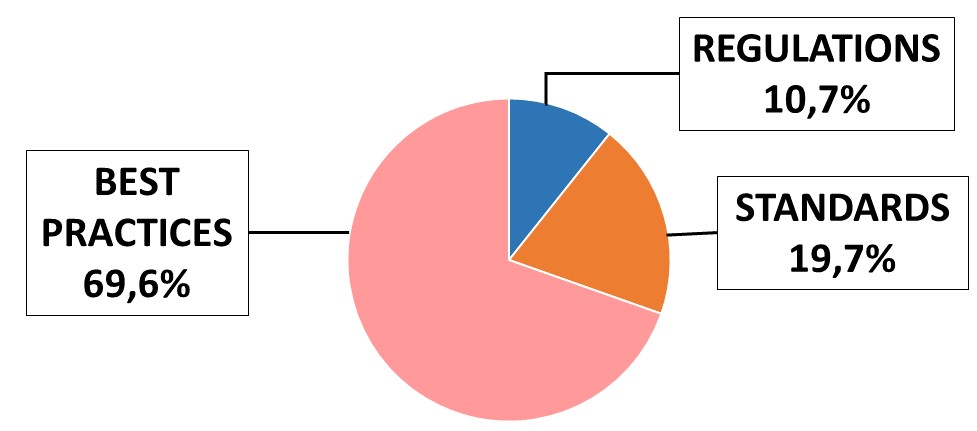}
\caption{Documents Analysis: excerpts distribution}
\label{fig:AnalysisDistribution}
\end{figure}


Based on the analysis of the 49 documents collected, we extracted approximately \emph{2,800} excerpts distributed as depicted in Figure~\ref {fig:AnalysisDistribution}.

\subsection{Documents Mapping}
\label{subsec:docMapping}

The excerpts identified in the previous step are listed in a table in their original form.
To systematize them, we choose the \textbf{NIST Cybersecurity Framework v1.1} that provides a common ground and standard terminology for cybersecurity functionalities.
However, since several key excerpts refer to data security and privacy, it was necessary to extend it. We leveraged the Italian Cybersecurity Framework~\cite{angelini2020italian,angelini2017crumbs}, retro-compatible with the NIST framework, that includes categories and subcategories dedicated to data protection.
Each excerpt has been accurately assessed for its semantic content and linked to one or more subcategories of the framework.\\
For each excerpt, the \emph{Function} it belongs to is first determined, followed by the assumed \emph{Category} and then the appropriate \emph{Subcategory}. An example of mapping is shown in Table~\ref {Table-mapping}.\\
This step mitigates \emph{issue 3} by identifying, quantifying, and resolving overlap.\\

\begin{figure}
\begin{minipage}[t]{0.4\textwidth}
\centering
\includegraphics[width=\textwidth]{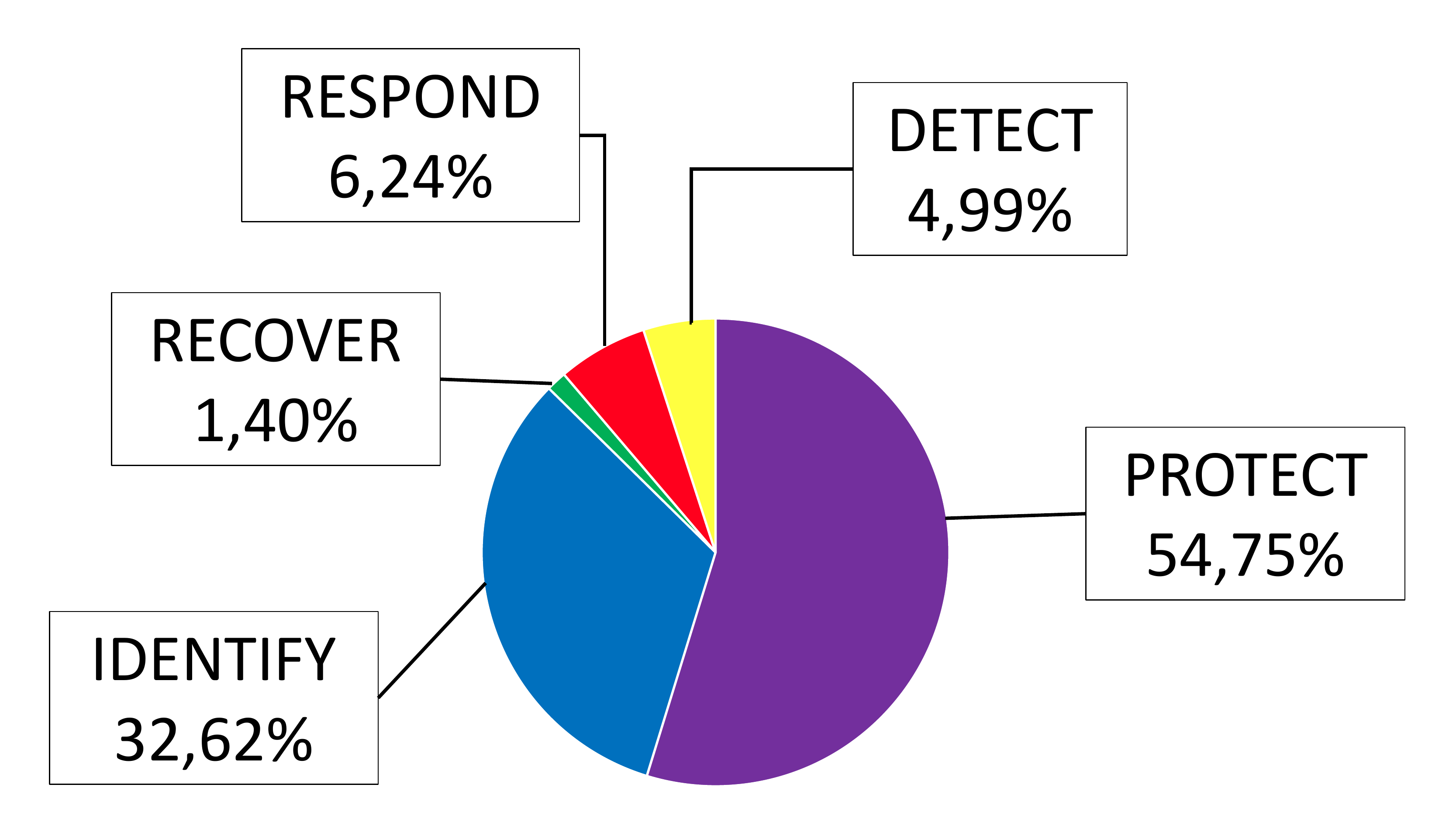}
\caption{Mapping (Functions)}
\label{fig:Functions}
\end{minipage}
\ \hspace{2mm} \hspace{3mm} \
\begin{minipage}[t]{0.5\textwidth}
\centering
\includegraphics[width=\textwidth]{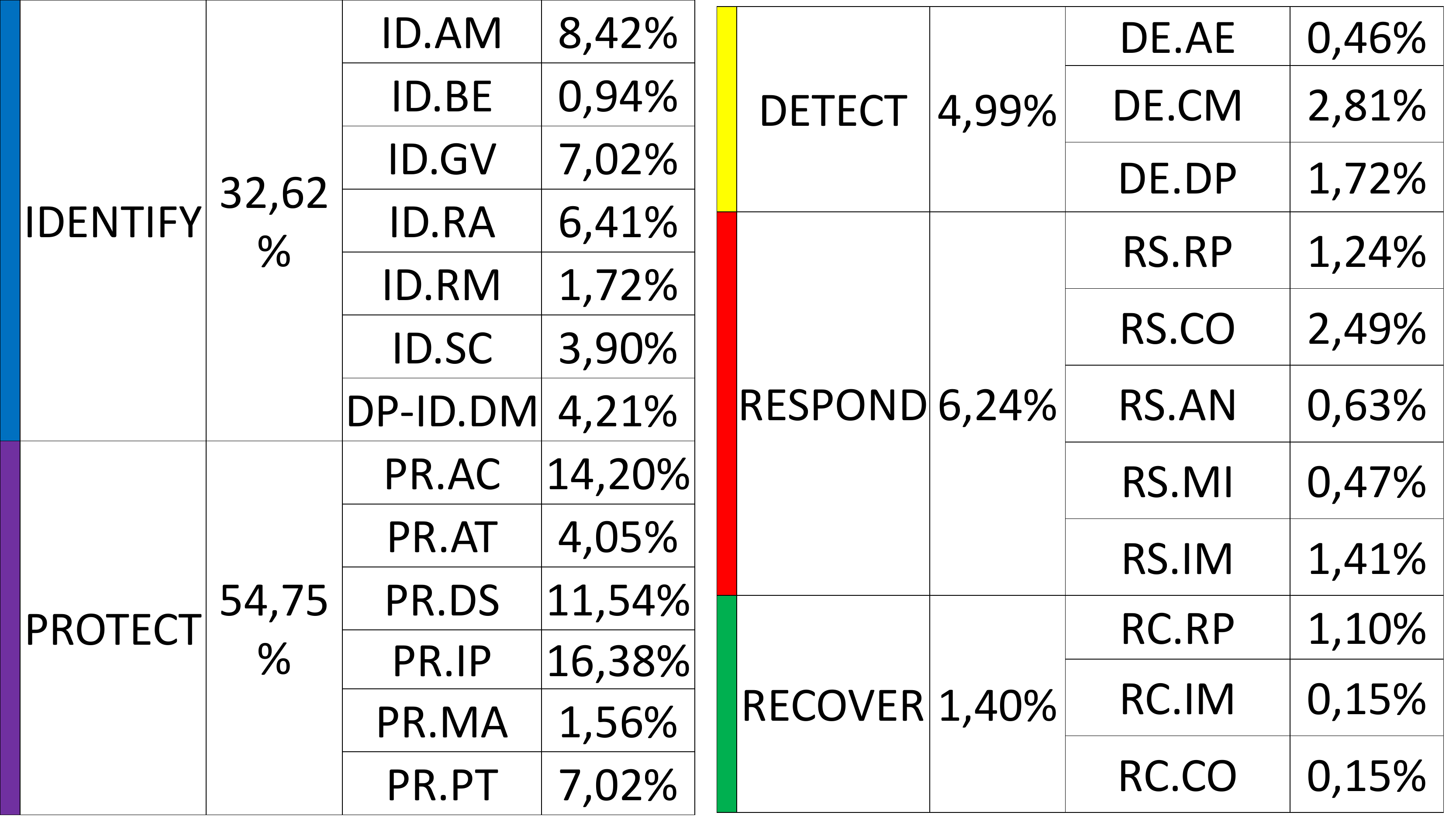}
\caption{Mapping (Categories)}
 \label{fig:Category}
\end{minipage}
\end{figure}

\medskip
\noindent
\textbf{Results.}
The 2,800 excerpts have been mapped mostly in the \emph{Protection} and \emph{Identify} functions. Very few excerpts address \emph{Respond}, \emph{Detect}, and \emph{Recover} functions, as visible in Figures~\ref {fig:Functions}.

\subsection{Controls Definition}
\label{subsec:controlsDefinition}

In this step, the excerpts previously mapped are refined and modeled as cybersecurity controls. It is necessary to refine the excerpts to be syntactically uniform because they were retrieved from documents of various types, origins, and writing styles.
For example, Best Practices have a purely technical nature and are made by sentences more direct and concise. In contrast, Regulations have a syntax typical of the legal world and are therefore made by sentences more discursive.\\
To define the controls and get a consistent and similar structure, \emph{three key constraints} were enforced during their definition:
\begin{enumerate}
    \item \textbf{Self-contained}: the control contains every element that is essential for its semantic completeness; 
    \item \textbf{Homogeneous}: the control faithfully complies with the semantics of the excerpt; 
    \item \textbf{Verifiable}: an application of the control must be verifiable through a well-defined quantitative or qualitative approach.
\end{enumerate}
A unique identifier then enumerates each control to retain its traceability.
By analyzing each excerpt in Table~\ref {Table-ControlsDef} and applying the constraints, one or more controls have been defined.
For example, a thorough semantic analysis of the excerpt \emph{E2} led to the definition of three controls: \emph{ID.GV-2-01} and \emph{ID.GV-2-02} directly derived from the original text while \emph{ID.AM- 6} has been added as an implicit requirement deriving from the former controls.

\begin{table}[!ht]
\small
    \centering
    \caption{Excerpts ext. \& Controls Definition Document \cite{ENISAsec} \emph{Recommendation 4 }}\label{Table-ControlsDef}
\begin{tabular}{ |p{5 cm}|p{2cm}|p{5 cm}|  }
 \hline
    \textbf{Excerpt Detail} & \textbf{\centering{Subcategory}} & \textbf{Control Definition} \\
 \hline
 E1: Member States should develop incident response mechanisms to efficiently bring together the healthcare organizations with the national cyber security competent centers. & \centering{PR.IP-9}   &  PR.IP-9-01: Healthcare organizations develop incident response mechanisms to bring together with the national cybersecurity competent centers  \\
\hline
E2: An eHealth-focused Computer Emergency Response Team should be created, which could potentially collaborate with the national CERT on incident handling. Feedback directly to the eHealth service users (e.g., clinicians) is extremely important for their continued engagement. & \centering{ID.GV-2; ID.AM- 6}  &  ID.GV-2-01: A Computer Emergency Response Team (CERT) has been created focused on eHealth.

ID.GV-2-02: The CERT collaborates with the national CERT on incident handling.

ID.AM-6-01: Roles and responsibilities are defined within the Computer Emergency Response Team \\

\hline
 E3: In terms of eHealth incident handling and hazard control, further steps need to be taken: Systems for reporting and analyzing incidents both locally and nationally. & \centering{RS.AN-5} &  
 RS.AN-5-01: Systems for reporting and analyzing incidents both locally and nationally have been implemented \\
 \hline
\end{tabular}
 \label{Table-mapping}
\end{table}
This step mitigates \emph{issues 1 and 2} by uniforming the contents and supporting the implementation of technical security measures.

\medskip
\noindent
\textbf{Results.}
At the end of this step, the approximately 2800 sentences extracted from the previous phase led to the definition of approximately \textbf{3,320 controls}. \\
The control definition's 15\% increase over the sentences extracted confirms the heterogeneity and fragmentation of the excerpt's content. 
The distribution of controls is uniform among the framework's categories (see Table~\ref {Table-ControlsDef}).
\section{Findings}
\label{sec:findings}
\medskip
\noindent
\textbf{Cybersecurity Controls Coverage.}
The first finding, depicted in Fig~\ref {fig:AnalysisDistribution}, is the large gap among the number of relevant excerpts extracted from Regulations and Standards compared to Best Practices.\\
The gap is mostly due to the nature of the documents themselves: \emph{Regulations} have the lowest percentage of extracted excerpts (10,7\%) because they are mostly discursive and do not address technological or procedural security measures, only stating general goals; \emph{Best Practices}, on the other hand, have the highest percentage of excerpts (69,60\%), since they are intended to serve as guidance for deploying cybersecurity measures, and therefore feature more technical and in-depth cybersecurity controls (\emph{Finding 1}).\\
While Healthcare organizations experienced significant security incidents in recent years, with the majority of them caused by either phishing or ransomware attacks \cite{akinsanya2019current}, there is still a lack of focus on how to address such threats. This is evidenced by the very low percentages of controls mapped on \emph{Detect} (4,99\%), \emph{Respond} (6,24\%), and \emph{Recovery} (1,40\%) functions, as shown in Figure~\ref {fig:Category}.
As a result, the documents focus mainly on the identification of cybersecurity perimeter and assets protection, with \emph{Protect} (54,75\%) and \emph{Identify} (32,62\%) being the most covered functions, rather than the detection and management of cybersecurity incidents during and after their deployment (\emph{Finding 2}).

\begin{figure}[!h]
\centering
\includegraphics[width=\textwidth]{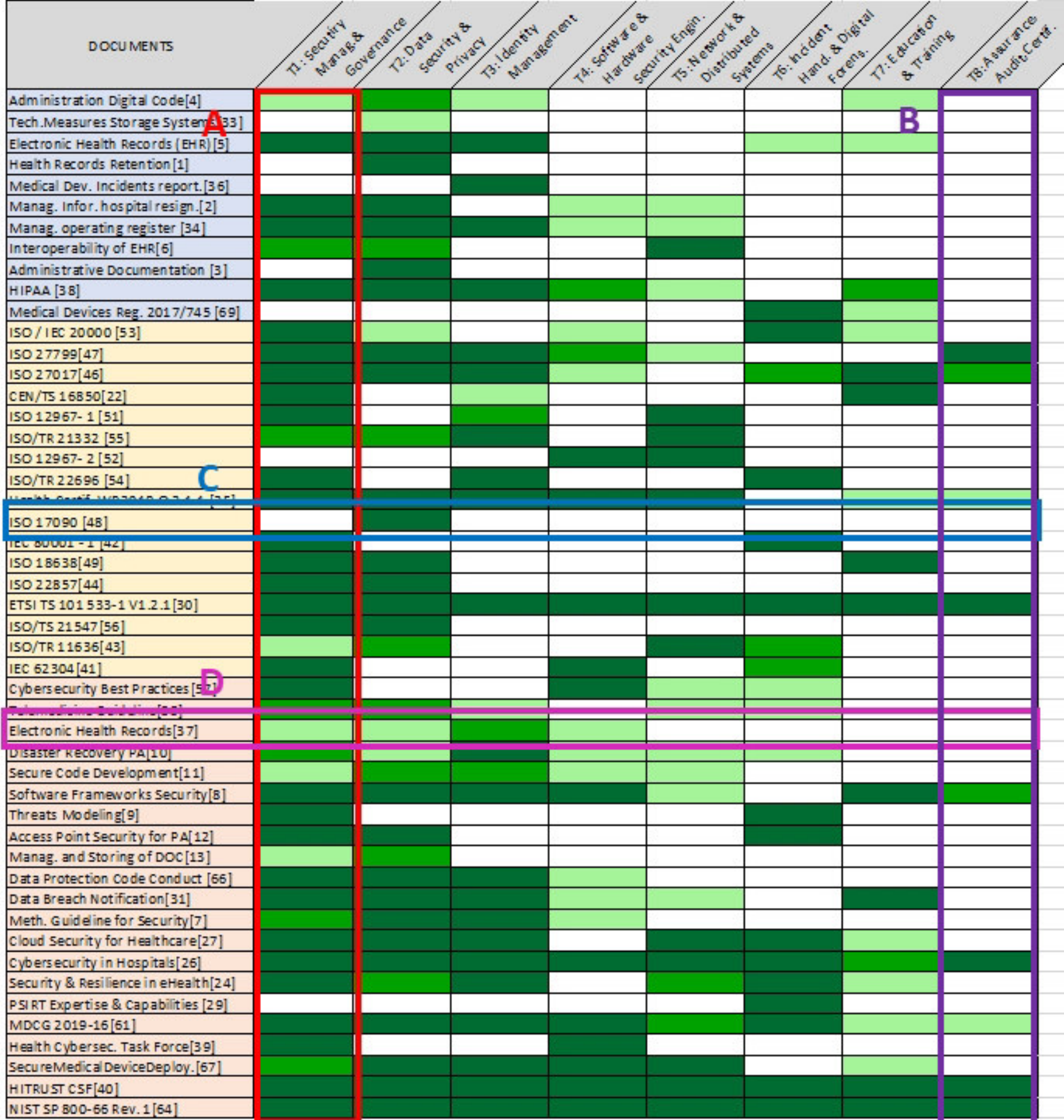}
\caption{Mapping Taxonomy}
\label{fig:taxonomy}
\end{figure}

\medskip
\noindent
\textbf{Cybersecurity Topics Coverage.}
The previous findings, 
were derived using the NIST Cybersecurity Framework. As it is a mostly operative framework, it gave us an idea of the less covered actions.
To provide additional insights focused on evaluating the coverage of key cybersecurity and data protection areas, we used a second  taxonomy suggested by the Report \emph{A Proposal for a European Cybersecurity Taxonomy} ~\cite{nai2019proposal}, issued by the European Commission.
We selected the most pertinent topics for the healthcare sector (eight) and used a three-stage approach to evaluate the coverage level for each topic. Firstly, the team assigned each subcategory of the NIST Framework to one of the taxonomy's security topics. Secondly, we counted the number of controls that fell into a specific subcategory for each document based on the mapping performed in the Control Definition Step. 
Thirdly, we developed three levels of coverage based on the number of occurrences: \emph{Low} if there were 1 to 3 controls that address the topic, \emph{Medium} (4 to 6 controls), and \emph{High} (more than 6 controls).
Thresholds were derived from a statistical analysis of the distribution of extracted controls per document per subcategory.
Figure~\ref{fig:taxonomy} shows the topics addressed and the level of coverage for each document gathered, ranging from dark green for high coverage to white for no coverage.
Using this approach, we provide an overview of documents coverage of cybersecurity topics and derive several findings.

\begin{figure}[h]
\centering \includegraphics[width=0.8\textwidth]{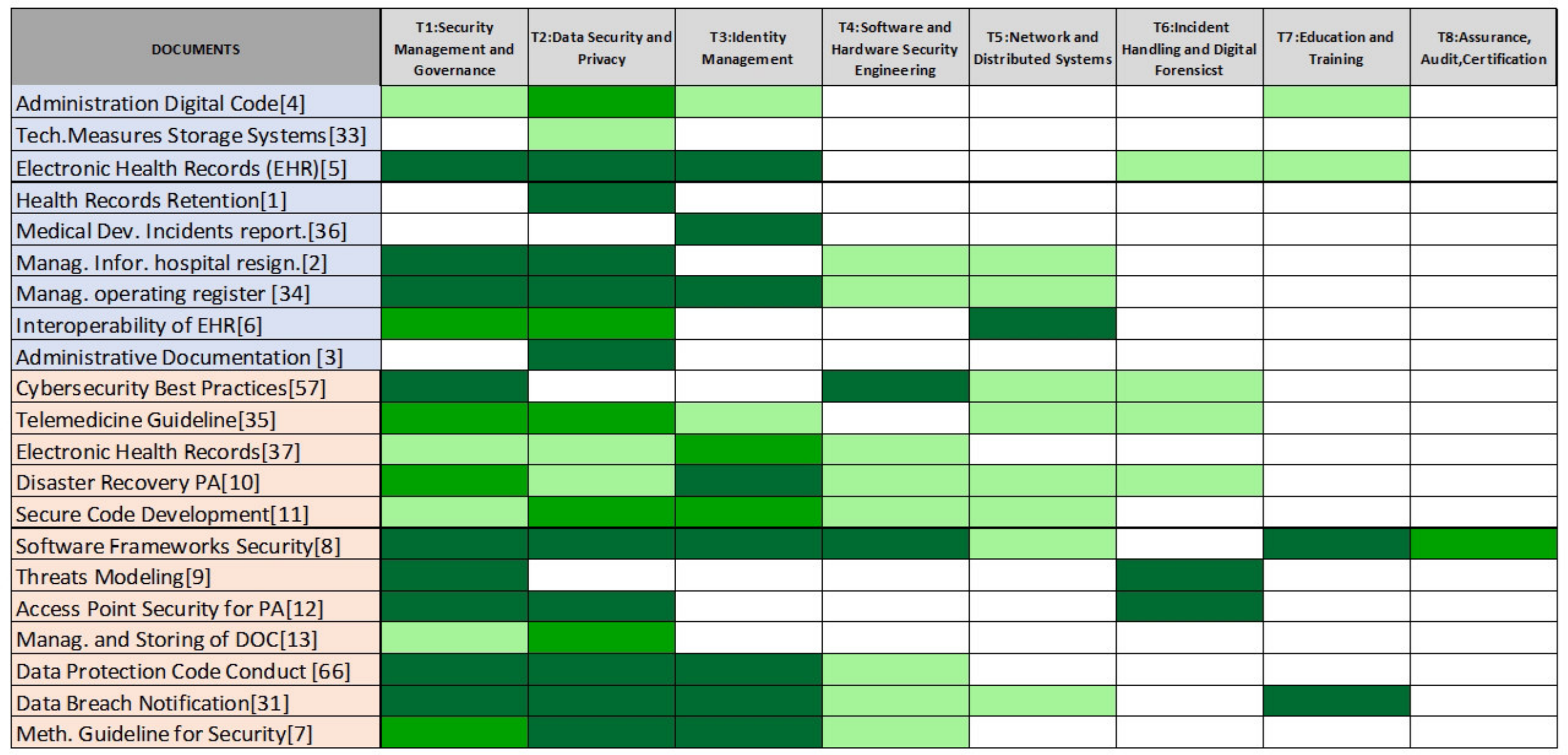}
\caption{National Mapping Taxonomy}
\label{fig:national_taxonomy}
\end{figure}

\emph{T1: Security Management and Governance} (Figure~\ref{fig:taxonomy} column A) is the most addressed topic (around 85\%) demonstrating the high regard by the documents collected. Moreover, the security measures are addressed by many publications issued by different sources, implying that the topic's contents are heavily overlapping (\emph{Finding 3}). In addition, many documents go deeper in their analysis (deep green color), and as a result, it could prove challenging to homogenize the security measures extracted.
Unlike T1, \emph{T8: Assurance, Audit, Certif.} (Figure~\ref{fig:taxonomy} column B) is the least addressed topic, with some scattered contributions from standards and best practices, even if focused.
Surprisingly, regulations do not provide cybersecurity controls in this area. Similar considerations can be made for topics \emph{T6: Incident Handling \& Digital Forens.} and \emph{T7: Education \& Training}. T6 presents a shallower coverage than T7, raising the possibility that the resulting security measures could be incomplete (\emph{Finding 4}).\\
Analyzing Figure~\ref{fig:taxonomy} from a document-based perspective, the coverage area is determined by the document typology: Regulations and Standards are more focused on specific topics, leaving others completely or partially uncovered, while Best Practices are broad and cross-topics. For instance, the standard \emph{ISO 17090}~\cite{17090} (row C) focuses on a single topic (T2), analyzing it in depth and providing specific security measures.
On the other hand, the \emph{National Best Practice for Electronic Health Record}~\cite{FSElinee} (row D) covers a wider range of topics, where most of them are only given a shallow level of analysis, implying that the contents are overly generic (\emph{Finding 5}).
An interesting analysis is to compare international and national coverage mappings (see Figure~\ref {fig:national_taxonomy}). 
Even among the National corpus of documents, the most popular topics are T1, \emph{T2: Data Security \& Privacy}, and \emph{T3: Identity Management}. It indicates that the national context tends to mirror the trend of international publications on these topics.
On the contrary, the remaining topics result less covered, both in terms of the number of controls and depth of analysis. Critical topics are T7 and T8, with the first addressed in less than 20\% of the documents and the second addressed by only one best practice (\emph{Finding 6}). Notice that there is a lack of standards in the list of national documents since all standards gathered during the collection step are issued by international entities.

\begin{figure}
\begin{minipage}[t]{0.45\textwidth}
\centering
\includegraphics[width=\textwidth]{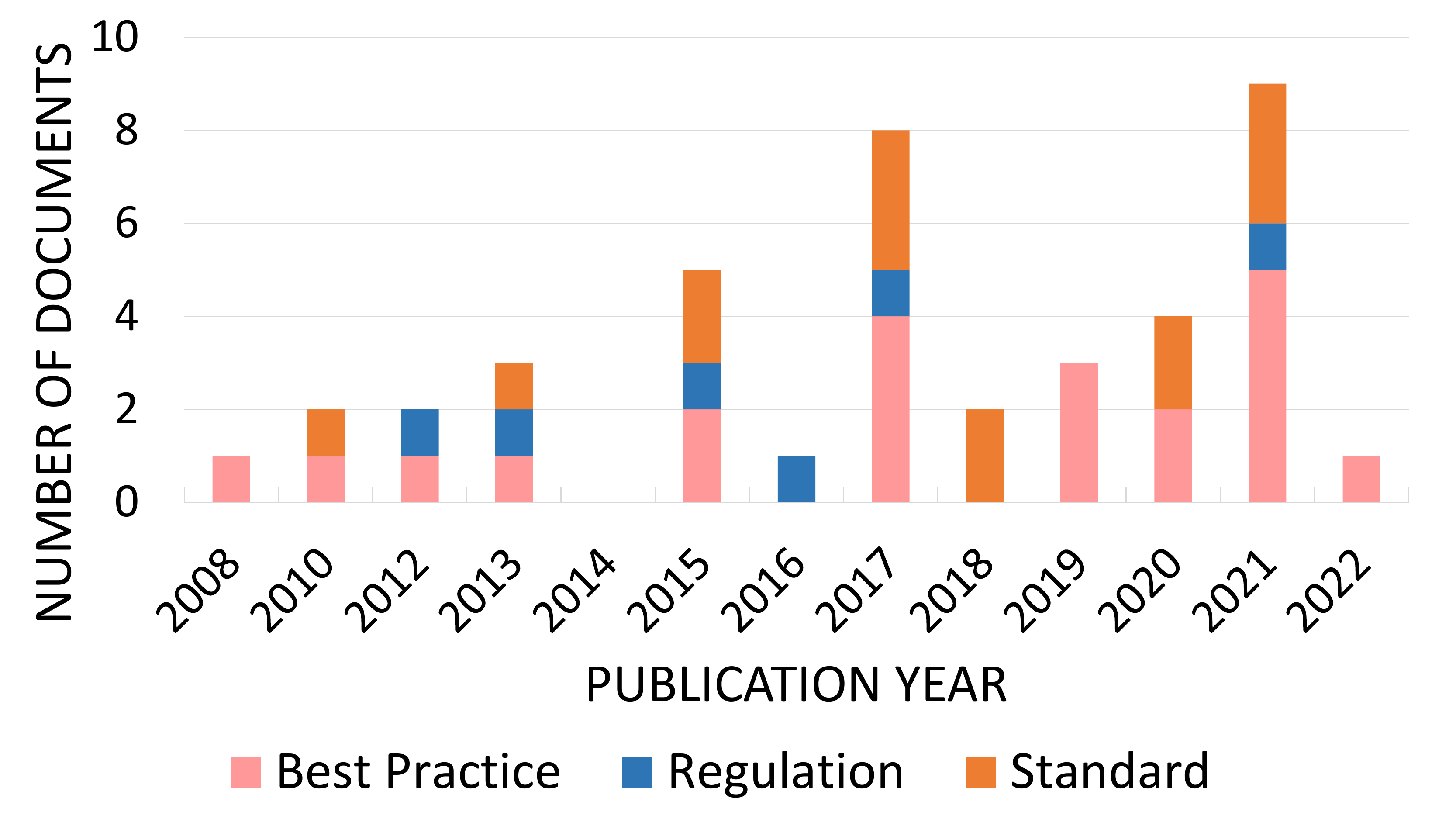}
\caption{Analysis by document type}
\label{fig:PubTimeline-a}
\end{minipage}
\ \hspace{2mm} \hspace{3mm} \
\begin{minipage}[t]{0.45\textwidth}
\centering
\includegraphics[width=\textwidth]{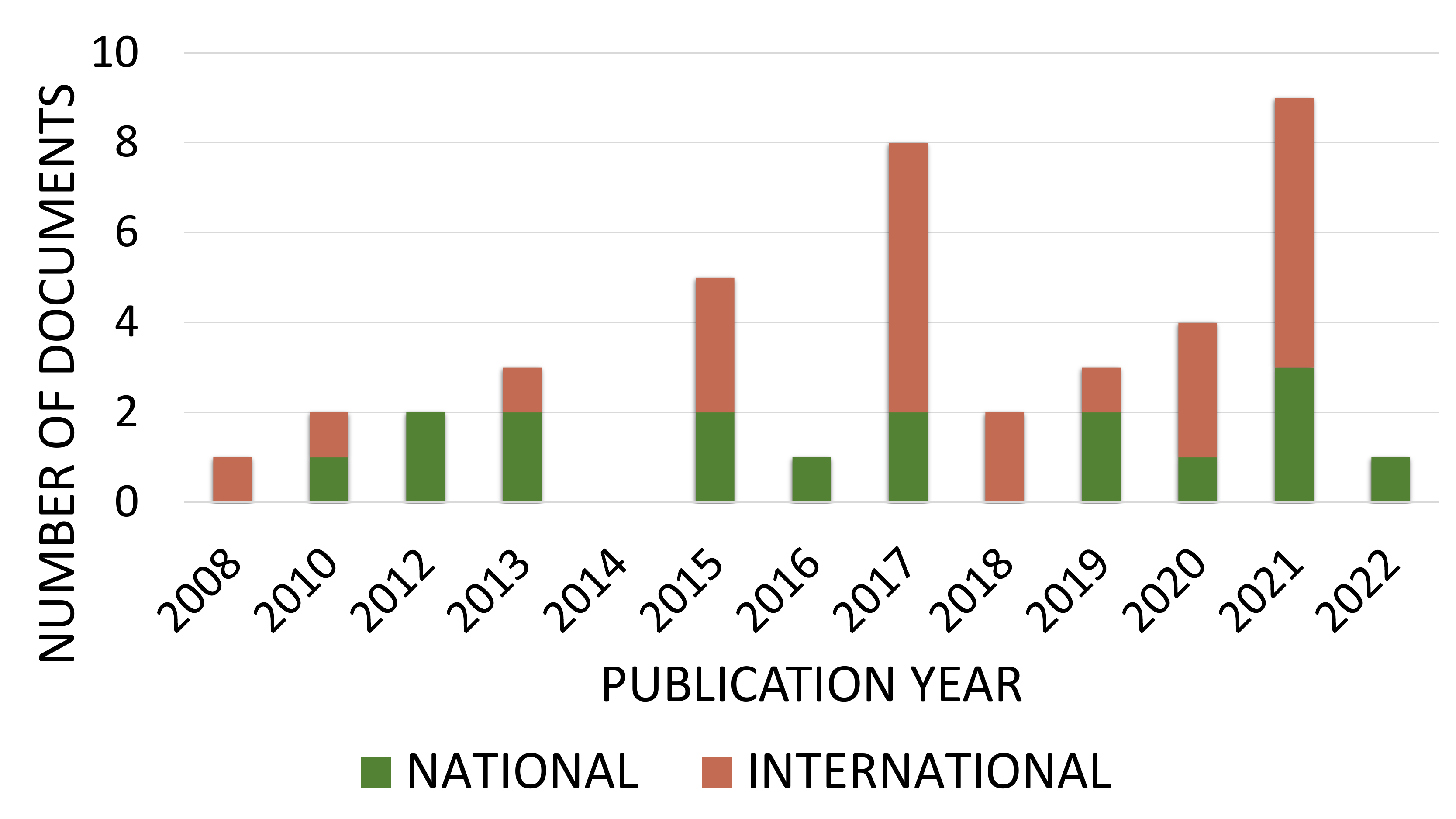}
\caption{National vs International}
 \label{fig:PubTimeline-b}
\end{minipage}
\end{figure}

\medskip
\noindent
\textbf{Temporal trends.}
The temporal analysis by date of publication confirms that cybersecurity is emerging as a top priority in the healthcare sector, with a steady increase in the pace of cybersecurity regulations, standards, and best practices publication since 2008.\\
As shown in Figure~\ref {fig:PubTimeline-a} there is a peak of publications in 2017 which may be related to the 2016 Hollywood Presbyterian ransomware attack (the first highly publicized cyberattack incident against a hospital)~\cite{argaw2019state} and a second peak in 2021, when, among others, regulations on medical devices, along with related guidelines and standards, have been published (\emph{Finding 7}).\\
Furthermore, Figure~\ref {fig:PubTimeline-b} illustrates that national publications tend to follow a similar trend, indicating that national authorities are attempting to keep up and align national regulations with international ones~(\emph{Finding 8}).

\medskip
\noindent
\textbf{Actors.}
A healthcare system is an organization of people, institutions, and resources that delivers services for the population.
We modeled, referring to literature, the healthcare sector as composed of five main providers, sorted by descending size:
\begin{itemize}
    \item \emph{Hospital}: an institution that provides diagnoses of disease, medical and surgical treatments, and nursing care for sick or injured people;
    \item \emph{Private Structure} (e.g., Care Homes,  Diagnostic Centers, etc. ): structure that performs several health services but cannot perform hospitalizations;
    \item \emph{Local Sanitary Unit}: the integrated primary health care public service covering a well-defined population;
    \item \emph{Clinical laboratory}: healthcare facility providing a wide range of laboratory procedures for diagnosis and treatment;
    \item \emph{Medical practitioner}: a self-employed or publicly employed health professional who works independently.
\end{itemize}

\begin{figure}[ht]
\centering \includegraphics[width=\textwidth]{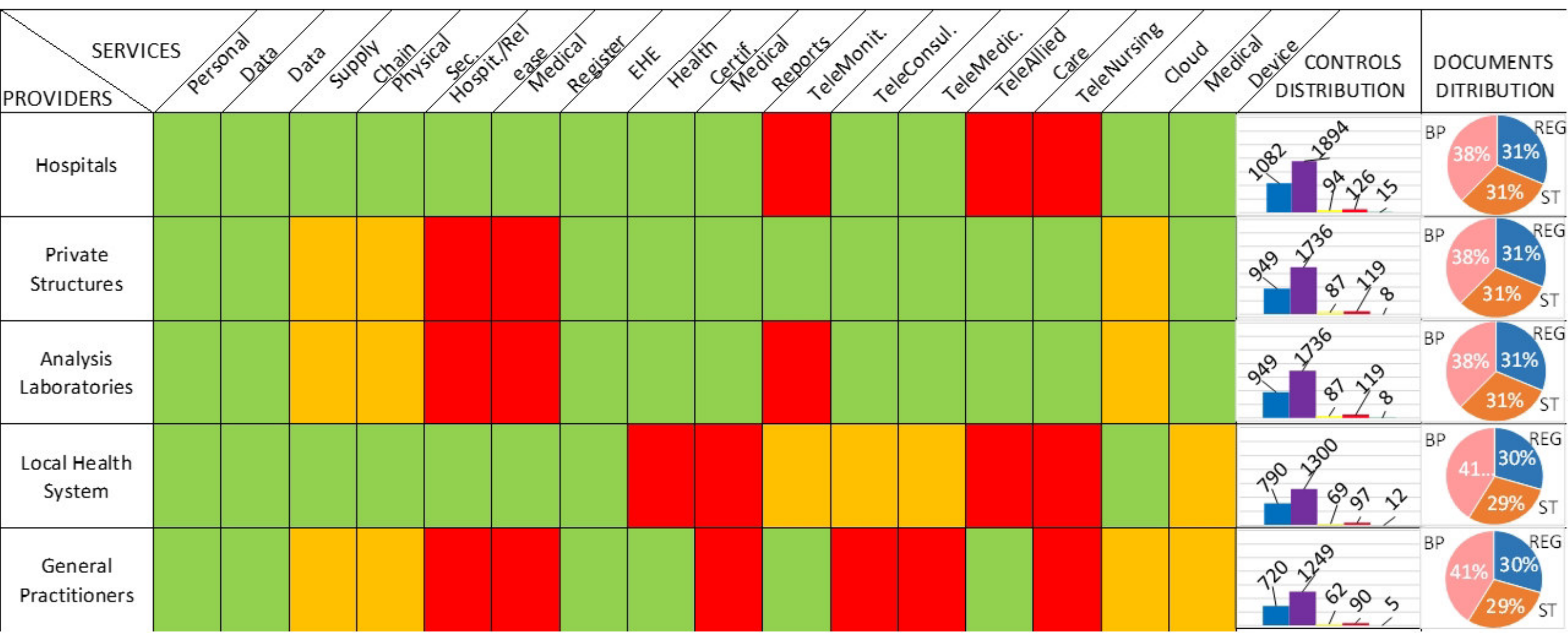}
\caption{Healthcare Providers and Services}
\label{fig:provserv}
\end{figure}

For each provider, we defined the delivered services classified as primary (compulsory to provide) mapped in green, secondary (optional to provide) indicated in yellow, and services not provided indicated in red (see Figure~\ref {fig:provserv}).

Afterward, for each primary service, we analyze which cybersecurity controls, defined in the Controls Definitions Step, could be fitting for securing the service.\\
As a result, for each provider, we obtain the number of cybersecurity controls that should be coped with to improve the providers’ cybersecurity posture distributed by functions and originating sources type (see Figure~\ref{fig:provserv}).\\
Due to the fewer services offered, Medical Practitioners need to cope with less than 60\% controls compared to a hospital organization. Overall, Identify and Protect remain the most addressed functions, and there is a uniform distribution of controls derived from Regulations, Standards, and Best Practices (\emph{Finding 9}). We notice that the number of controls to cope remains high, disregarding the target actor. More effort should be put in place to streamline their implementation, considering their priority and the security of secondary services.

\section{Conclusions}
\label{sec:conclusions}
This paper systematized healthcare sector cybersecurity and data protection regulations, standards, and best practices, analyzing 49 documents and categorizing them using the NIST Framework. This resulted in 3200 security controls and nine findings, including that best practices present more technical controls than Regulations. We found an uneven distribution of controls for cybersecurity and data protection topics, particularly in the areas of Detect, Respond, and Recover. Future plans include updating the systematization with new documents, like NIS2, and utilizing the controls for cyber-posture assessments.

\nocite{*}
\bibliographystyle{splncs04}
\bibliography{biblio}

\clearpage
\appendix
\include{Sections/Appendix.tex}

\end{document}